\documentclass[english]{article}
\usepackage[T1]{fontenc}
\usepackage[latin9]{inputenc}
\usepackage{geometry}
\usepackage{bm}
\usepackage{amsmath}

\makeatletter
\usepackage{indentfirst}

\makeatother

\usepackage{babel}
\begin{document}
\title{\textbf{\Large{}Clifford-algebraic formulation of nonlinear conformal
transformations in electrodynamics}}
\author{{\normalsize{}Leehwa Yeh}\thanks{Electronic mail: yehleehwa@gmail.com}}
\date{\textit{\small{}Shing-Tung Yau Center, National Yang Ming Chiao Tung
University, Hsinchu 30010, Taiwan}}

\maketitle
\noindent We derive a set of Clifford-algebraic formulas for two major
nonlinear conformal transformations of the physical quantities related
to Maxwell's equations. The superiority of these formulas over their
vector-tensorial counterparts are demonstrated through three illustrative
examples.\\


\subsection*{I. INTRODUCTION}

Maxwell\textquoteright s equations are renowned for being form-invariant
under Lorentz transformation, which is a linear transformation of
spacetime. As shown by Cunningham and Bateman back in 1909 \cite{key-1,key-2},
this invariance can be extended to conformal transformation which
is nonlinear in general. 

While the manifestations of conformal transformation in classical
electrodynamics have been extensively explored \cite{key-3}, the
inherent complexity of the mathematical formulation continues to pose
challenges for both novices and experienced researchers in this field.
This complexity, in our opinion, arises mainly from the prevalent
use of vectors and tensors in electrodynamics as well as special relativity.
Although the advantage of invoking Clifford algebra (also known as
geometric algebra) as a substitute is recognized by some authors \cite{key-4,key-5,key-6},
there has been little research on applying Clifford algebra to conformal
transformation strictly in four-dimensional spacetime.

Inspired by the Clifford-algebraic versions of the Lorentz transformations
for four-vectors and electromagnetic tensor, we set a goal to derive
the analogues of conformal inversion, which is the simplest nonlinear
conformal transformation. Through a series of attempts, we achieved
a set of concise formulas in covariant as well as non-covariant Clifford-algebraic
formulations. Building on this foundation, we further generalized
the formulas to apply to the most important nonlinear conformal transformation,
the special conformal transformation (SCT). As expected, these formulas
turned out to exhibit superior simplicity and elegance compared to
their vector-tensorial counterparts.

The organization of this paper is as follows: In the next section
we introduce the necessary background knowledge, then we derive the
said formulas step by step in Sec. III. In order to demonstrate the
superiority of these formulas, we work on three practical examples
in Sec. IV. Finally in Sec. V we enumerate the advantages of these
newly-found formulas. Two appendices are provided at the end of this
paper to enhance readers' comprehension of this research. 

\section*{II. PRELIMINARIES}

\subsection*{A. Conventions and notations}

Throughout this paper $c=1$, and all quantities are nondimensionalized.

Euclidean space is defined as a three-dimensional dimensionless space
with the coordinates $(x,y,z)$ and metric $\text{diag}(1,1,1)$.
The corresponding orthonormal basis is denoted by $\{\hat{x},\hat{y},\hat{z}\}$,
and the position vector by
\begin{equation}
\vec{r}=x\hat{x}+y\hat{y}+z\hat{z}.
\end{equation}
A Euclidean vector is denoted by 
\begin{equation}
\vec{U}=U_{x}\hat{x}+U_{y}\hat{y}+U_{z}\hat{z},
\end{equation}
and the scalar product of two Euclidean vectors by
\begin{equation}
\vec{U}\cdot\vec{V}=U_{x}V_{x}+U_{y}V_{y}+U_{z}V_{z}.
\end{equation}

Minkowski space is defined as a (1+3)-dimensional dimensionless space
with the coordinates $\text{x}^{\alpha}=(t,x,y,z)$ and metric $\text{diag}(1,-1,-1,-1)=:\eta_{\alpha\beta}$,
where the Greek indices run from 0 to 3. The corresponding orthonormal
basis is denoted by $\{\mathsf{e_{\alpha}}\}$, and the position four-vector
by 
\begin{equation}
\text{x}=\text{x}^{\alpha}\mathsf{e_{\alpha}}\label{4}
\end{equation}
under the summation convention. Other four-vectors follow the same
convention, for example,
\begin{equation}
A=A^{\alpha}\mathsf{e_{\alpha}}\label{5}
\end{equation}
 means the contravariant components of $A$ are 
\begin{equation}
A^{\alpha}=(A^{0},A^{1},A^{2},A^{3})=(A_{0},-A_{1},-A_{2},-A_{3}).
\end{equation}
The scalar product of two four-vectors is denoted by
\begin{equation}
A\cdot J=\eta_{\alpha\beta}A^{\alpha}J^{\beta}=A^{\alpha}J_{\alpha}.
\end{equation}

We always use $\text{x}'^{\mu}=(t',x',y',z')$ for the new coordinates
after conformal inversion, and $\text{x}''^{\mu}=(t'',x'',y'',z'')$
for that after special conformal transformation. For a generic conformal
transformation, the new coordinates are denoted by $\tilde{\text{x}}^{\mu}=(\tilde{t},\tilde{x},\tilde{y},\tilde{z})$,
hence both $x'^{\mu}$ and $x''^{\mu}$ are special cases of $\tilde{x}^{\mu}$.
Functions of these coordinates follow the same convention, for example,
$\vec{E}'(t',x',y',z')$ denotes the transformed electric field after
conformal inversion.

\subsection*{B. Conformal transformation }

The conformal transformation 
\begin{equation}
\tilde{\text{x}}^{\mu}=\mathcal{C}^{\mu}(\text{x})\label{8}
\end{equation}
is defined as a one-to-one relation in Minkowski space that satisfies
\begin{equation}
\Lambda{}^{2}\frac{\partial\tilde{\text{x}}^{\mu}}{\partial\text{x}^{\alpha}}\frac{\partial\tilde{\text{x}}^{\nu}}{\partial\text{x}^{\beta}}\eta_{\mu\nu}=\eta_{\alpha\beta}\text{ with }\Lambda>0.\label{9}
\end{equation}
It is obvious that $\Lambda$ is related to the Jacobian determinant
of this transformation, i.e.,
\begin{equation}
\Lambda=\bigg|\det\bigg[\frac{\partial\tilde{\text{x}}}{\partial\text{x}}\bigg]\bigg|^{-\frac{1}{4}}.
\end{equation}

There exist two interpretations for Eq. (\ref{8}) that correspond
to the active and the passive transformations. In this paper we adopt
the former, i.e., interpreting Eq. (\ref{8}) as mapping each spacetime
point and its contents to another point. The results we obtain are
also valid for the so-called conformal coordinate transformation \cite{key-7}
which is the passive conformal transformation plus a scale transformation,
a detailed discussion can be found in \cite{key-8}.

It can be proved that every conformal transformation is a composition
of four fundamental ones \cite{key-9} : ($\bm{\mathit{1}}$) Four-dimensional
dilation 
\begin{equation}
\tilde{\text{x}}=\lambda^{-1}\text{x}\iff\tilde{\text{x}}^{\mu}=\lambda^{-1}\text{x}^{\mu},\label{11}
\end{equation}
where $\lambda$ is a positive constant; ($\bm{\mathit{2}}$) Four-dimensional
translation
\begin{equation}
\tilde{\text{x}}=\text{x}+\text{b}\iff\tilde{\text{x}}^{\mu}=\text{x}^{\mu}+\text{b}^{\mu},\label{12}
\end{equation}
where $\text{b}=\text{b}^{\mu}\mathsf{e}_{\mu}$ is a constant four-vector;
($\bm{\mathit{3}}$) Lorentz transformation
\begin{equation}
\tilde{\text{x}}^{\mu}=\mathcal{L}_{\alpha}^{\mu}\text{x}^{\alpha}\text{ with }\mathcal{L}_{\alpha}^{\mu}\mathcal{L}_{\beta}^{\nu}\eta_{\mu\nu}=\eta_{\alpha\beta},\label{13}
\end{equation}
where $\mathcal{L}$ is a constant matrix, and ($\bm{\mathit{4}}$)
Special conformal transformation (abbreviated as SCT henceforward)
which is defined in Eq. (\ref{19}). The degrees of freedom of these
transformations are one, four, six, and four respectively.

\subsection*{C. Conformal inversion }

Examples of linear conformal transformation include Eqs. (\ref{11})$-$(\ref{13}).
As for the nonlinear ones, the simplest example is the conformal inversion
\begin{equation}
\text{x}'=\varepsilon\text{x}/\text{x}^{2}\iff\text{x}'^{\mu}=\varepsilon\text{x}^{\mu}/\text{x}^{2},\label{14}
\end{equation}
where $\varepsilon=1$ or $-1$, and 
\begin{equation}
\text{x}^{2}:=t^{2}-r^{2}\ne0,\label{15}
\end{equation}
 i.e., the light cone $t^{2}=r^{2}$ is excluded from the domain of
Eq. (\ref{14}). This transformation will be referred to as Inversion
from now on.

From the Jacobian matrix element 
\begin{equation}
\frac{\partial\text{x}'^{\mu}}{\partial\text{x}^{\alpha}}=\varepsilon\big(\text{x}^{2}\delta_{\alpha}^{\mu}-2\text{x}^{\mu}\text{x}_{\alpha}\big)\big/(\text{x}^{2})^{2},\label{16}
\end{equation}
we can prove that Inversion is a conformal transformation with 
\begin{equation}
\Lambda=|\text{x}^{2}|=|\text{x}'^{2}|^{-1}.\label{17}
\end{equation}

\subsection*{D. Special conformal transformation (SCT)}

As the only nonlinear fundamental conformal transformation, SCT is
made of three consecutive conformal transformations (Inversion$\longrightarrow$translation$\longrightarrow$Inversion):
\begin{align}
 & \text{x}'=\varepsilon\text{x}/\text{x}^{2}\ \text{with }\text{x}^{2}\ne0,\nonumber \\
 & \tilde{\text{x}}=\text{x}'+\varepsilon\text{a}\ \text{with }\text{a}^{2}\ne0,\label{18}\\
 & \text{x}''=\varepsilon\tilde{\text{x}}/\tilde{\text{x}}^{2}\ \text{with }\tilde{\text{x}}^{2}\ne0.\nonumber 
\end{align}
The composition of Eq. (\ref{18}) takes the form
\begin{equation}
\text{x}''=\Sigma^{-1}(\text{x}+\text{x}^{2}\text{a})\iff\text{x}''^{\mu}=\Sigma^{-1}\big(\text{x}^{\mu}+\text{x}^{2}\text{a}^{\mu}\big),\label{19}
\end{equation}
where 
\begin{align}
\Sigma: & =\text{x}^{2}\tilde{\text{x}}^{2}=(\text{x}'^{2}\text{x}''^{2})^{-1}\nonumber \\
 & =1+2\text{a}\cdot\text{x}+\text{a}^{2}\text{x}^{2}=\big(1-2\text{a}\cdot\text{x}''+\text{a}^{2}\text{x}''^{2}\big)^{-1}.\label{20}
\end{align}

It is easy to prove that Eqs. (\ref{18}) and (\ref{19}) are conformal
transformations with $\Lambda=|\Sigma|$, and $\Sigma=0$ is a light
cone whose vertex lies at $-\text{a}^{\mu}/\text{a}^{2}$ \cite{key-10}. 

\subsection*{E. Conformal transformations of electrodynamic quantities}

Under the active conformal transformation where the metric is fixed,
the transformation laws for the components of potential four-vector,
current density four-vector (``current four-vector\textquotedblright{}
for short henceforward), and electromagnetic tensor are as follows
\cite{key-8}.
\begin{align}
\tilde{A}^{\mu}(\tilde{\text{x}}) & =\vartheta\Lambda{}^{2}\frac{\partial\tilde{\text{x}}^{\mu}}{\partial\text{x}^{\alpha}}A^{\alpha}(\text{x}),\label{21}\\
\tilde{A}_{\mu}(\tilde{\text{x}}) & =\eta_{\mu\nu}\tilde{A}^{\nu}(\tilde{\text{x}})=\vartheta\frac{\partial\text{x}^{\alpha}}{\partial\tilde{\text{x}}^{\mu}}A_{\alpha}(\text{x}),\label{22}\\
\tilde{J}^{\mu}(\tilde{\text{x}}) & =\vartheta\Lambda{}^{4}\frac{\partial\tilde{\text{x}}^{\mu}}{\partial\text{x}^{\alpha}}J^{\alpha}(\text{x}),\label{23}\\
\tilde{J}_{\mu}(\tilde{\text{x}}) & =\eta_{\mu\nu}\tilde{J}^{\nu}(\tilde{\text{x}})=\vartheta\Lambda{}^{2}\frac{\partial\text{x}^{\alpha}}{\partial\tilde{\text{x}}^{\mu}}J_{\alpha}(\text{x}),\label{24}\\
\tilde{F}^{\mu\nu}(\tilde{\text{x}}) & =\vartheta\Lambda{}^{4}\frac{\partial\tilde{\text{x}}^{\mu}}{\partial\text{x}^{\alpha}}\frac{\partial\tilde{\text{x}}^{\nu}}{\partial\text{x}^{\beta}}F^{\alpha\beta}(\text{x}),\label{25}\\
\tilde{F}_{\mu\nu}(\tilde{\text{x}}) & =\eta_{\mu\rho}\eta_{\nu\sigma}\tilde{F}^{\rho\sigma}(\tilde{\text{x}})=\vartheta\frac{\partial\text{x}^{\alpha}}{\partial\tilde{\text{x}}^{\mu}}\frac{\partial\text{x}^{\beta}}{\partial\tilde{\text{x}}^{\nu}}F_{\alpha\beta}(\text{x}),\label{26}
\end{align}
where 
\begin{equation}
\ensuremath{\text{\ensuremath{\vartheta}}:=}\ensuremath{\text{sgn}}\bigg(\frac{\partial\tilde{t}}{\partial t}\bigg).
\end{equation}
For Inversion, we can derive
\begin{equation}
\vartheta=-\varepsilon\label{28}
\end{equation}
from Eq. (\ref{16}), and it follows that $\vartheta=(-\varepsilon)^{2}=1$
for SCT. 

\section*{III. DERIVATIONS OF THE FORMULAS }

\subsection*{A. Review of Clifford algebra $C\!\ell_{1,3}$}

Clifford algebra $C\!\ell_{1,3}$ of Minkowski space is defined as
a 16-dimensional vector space with the basis
\begin{equation}
\{1,\mathsf{e}_{\alpha},\mathsf{e}_{\alpha}\mathsf{e}_{\beta},\mathsf{e}_{\alpha}\mathsf{e}_{\beta}\mathsf{e}_{\gamma},\mathsf{e}_{0}\mathsf{e}_{1}\mathsf{e}_{2}\mathsf{e}_{3}\},
\end{equation}
where the associative product between $\mathsf{e}_{\alpha}\!$'s is
called geometric product and is defined by
\begin{equation}
\mathsf{e}_{\alpha}\mathsf{e}_{\beta}+\mathsf{e}_{\beta}\mathsf{e}_{\alpha}=2\eta_{\alpha\beta}.\label{30}
\end{equation}

Once the geometric product is introduced, $\text{x}^{2}$ in Eq. (\ref{15})
is no longer a symbol. It is straightforward to derive $\text{x}\text{y}+\text{y}\text{x}=2\text{x}\cdot\text{y}$
from Eqs. (\ref{4}) and (\ref{30}), and it follows that
\begin{equation}
\text{x}^{2}=\text{x}\text{x}=\text{x}\cdot\text{x}=t^{2}-r^{2}.\label{31}
\end{equation}
Accordingly, the left hand side of Eq. (\ref{19}) can be written
as
\begin{equation}
\text{x}''=\Sigma^{-1}(1+\text{a}\text{x})\text{x}=\Sigma^{-1}\text{x}(1+\text{x}\text{a}),\label{32}
\end{equation}
where
\begin{align}
\Sigma & =1+2\text{a}\cdot\text{x}+\text{a}^{2}\text{x}^{2}=(1+\text{a}\text{x})(1+\text{x}\text{a})=(1+\text{x}\text{a})(1+\text{a}\text{x})\nonumber \\
 & =\big(1-2\text{a}\cdot\text{x}''+\text{a}^{2}\text{x}''^{2}\big)^{-1}=\big[(1-\text{a}\text{x}'')(1-\text{x}''\text{a})\big]^{-1}=\big[(1-\text{x}''\text{a})(1-\text{a}\text{x}'')\big]^{-1}.\label{33}
\end{align}

When we use $C\!\ell_{1,3}$ as the mathematical tool for electrodynamics
\cite{key-4,key-5}, as long as magnetic monopole is not taken into
account, only two kinds of basis vectors of this algebra are needed,
i.e., $\{\mathsf{e}_{\alpha}\}$ for four-vectors, and $\{\mathsf{e}_{\alpha}\mathsf{e}_{\beta}\}$
for bivectors that correspond to anti-symmetric tensors of rank two.
We will discuss them separately in the following paragraphs.

(i) $\{\mathsf{e}_{\alpha}\}$ : Both of the potential and current
four-vectors take the form of Eq. (\ref{5}). Under four-dimensional
dilation Eq. (\ref{11}), the four-vector $W(\text{x})$ transforms
as

\begin{equation}
\tilde{W}(\tilde{\text{x}})=\lambda^{n}W(\text{x}),\label{34}
\end{equation}
where $n=1$ for potential four-vector and $n=3$ for current four-vector
according to scaling law; while under four-dimensional translation
Eq. (\ref{12}), the corresponding transformation is simply
\begin{equation}
\tilde{W}(\tilde{\text{x}})=W(\text{x}).\label{35}
\end{equation}

Involving no geometric product, Eqs. (\ref{34}) and (\ref{35}) are
the same as those in vector-tensorial formulation. On the other hand,
formulas of Lorentz transformations usually contain geometric products.
For example, parity transformation
\begin{equation}
\tilde{\text{x}}^{\mu}=(t,-x,-y,-z),\label{36}
\end{equation}
which is one of the simplest Lorentz transformations, leads to the
transformation law
\begin{equation}
\tilde{W}(\tilde{\text{x}})=\mathsf{e}_{0}W(\text{x})\mathsf{e}_{0},\label{37}
\end{equation}
which includes the Clifford-algebraic version of Eq. (\ref{36}),
$\tilde{\text{x}}=\mathsf{e}_{0}\text{x}\mathsf{e}_{0}$, as a special
case. 

Note that Eq. (\ref{34}) is compatible with Eqs. (\ref{21})$-$(\ref{24})
with $\vartheta=1$ and $\Lambda=\lambda,$ while both Eqs. (\ref{35})
and (\ref{37}) are compatible with Eqs. (\ref{21})$-$(\ref{24})
with $\vartheta=1$ and $\Lambda=1$.

For completeness, we list the Clifford-algebraic formulas for four
kinds of Lorentz transformations in Appendix II. 

(ii) $\{\mathsf{e}_{\alpha}\mathsf{e}_{\beta}\}$ : In electrodynamics,
the most important bivector is 
\begin{equation}
F=\tfrac{1}{2}F^{\alpha\beta}\mathsf{e}_{\alpha}\mathsf{e}_{\beta},\label{38}
\end{equation}
where
\begin{equation}
F^{\alpha\beta}=\begin{pmatrix}0 & -E_{x} & -E_{y} & -E_{z}\\
E_{x} & 0 & -B_{z} & B_{y}\\
E_{y} & B_{z} & 0 & -B_{x}\\
E_{z} & -B_{y} & B_{x} & 0
\end{pmatrix}
\end{equation}
is the contravariant electromagnetic tensor. $F$ in Eq. (\ref{38})
is called Faraday bivector, and its explicit expression is
\begin{equation}
F=E_{x}\mathsf{e}_{1}\mathsf{e}_{0}+E_{y}\mathsf{e}_{2}\mathsf{e}_{0}+E_{z}\mathsf{e}_{3}\mathsf{e}_{0}+B_{x}\mathsf{e}_{3}\mathsf{e}_{2}+B_{y}\mathsf{e}_{1}\mathsf{e}_{3}+B_{z}\mathsf{e}_{2}\mathsf{e}_{1}.
\end{equation}

Under four-dimensional dilation Eq. (\ref{11}), four-dimensional
translation Eq. (\ref{12}), and parity transformation Eq. (\ref{36}),
Faraday bivector transforms respectively as
\begin{align}
 & \tilde{F}(\tilde{\text{x}})=\lambda^{2}F(\text{x}),\label{41}\\
 & \tilde{F}(\tilde{\text{x}})=F(\text{x}),\label{42}\\
 & \tilde{F}(\tilde{\text{x}})=\mathsf{e}_{0}F(\text{x})\mathsf{e}_{0},\label{43}
\end{align}
where Eq. (\ref{41}) is compatible with Eqs. (\ref{25}) and (\ref{26})
with $\vartheta=1$ and $\Lambda=\lambda$, while both Eqs. (\ref{42})
and (\ref{43}) are compatible with Eqs. (\ref{25}) and (\ref{26})
with $\vartheta=1$ and $\Lambda=1$.

It is worthwhile to point out the similarity between Eqs. (\ref{37})
and (\ref{43}), which finds no analogue in vector-tensorial formulation. 

\subsection*{B. Formulas for Inversion }

In order to derive the formulas associated with Inversion that imitate
Eqs. (\ref{37}) and (\ref{43}), we first introduce an identity
\begin{equation}
\text{x}^{4}\frac{\partial\text{x}'^{\mu}}{\partial\text{x}^{\alpha}}\mathsf{e}_{\mu}=-\varepsilon\text{x}\mathsf{e}_{\alpha}\text{x},\label{44}
\end{equation}
which can be proved via Eq. (\ref{16}) and
\begin{equation}
\mathsf{e}_{\alpha}\text{x}+\text{x}\mathsf{e}_{\alpha}=2\text{x}_{\alpha},
\end{equation}
which is a derivative of Eq. (\ref{30}).

Using this identity in company with Eqs. (\ref{17}), (\ref{21}),
and (\ref{28}), we obtain the sought-for formula for potential four-vector,
\begin{equation}
A'=A'^{\mu}\mathsf{e}_{\mu}=(-\varepsilon)\text{x}^{4}\frac{\partial\text{x}'^{\mu}}{\partial\text{x}^{\alpha}}A^{\alpha}\mathsf{e}_{\mu}=\text{x}A\text{x},\label{46}
\end{equation}
where the arguments of $A'=A'(\text{x}')$ and $A=A(\text{x})$ are
related by Eq. (\ref{14}). 

The Inversion formula of Faraday bivector can be derived as follows:
According to Eqs. (\ref{17}), (\ref{25}), and (\ref{28}),
\begin{equation}
F'=\tfrac{1}{2}F'^{\mu\nu}\mathsf{e}_{\mu}\mathsf{e}_{\nu}=\tfrac{1}{2}(-\varepsilon)\text{x}^{8}\frac{\partial\text{x}'^{\mu}}{\partial\text{x}^{\alpha}}\frac{\partial\text{x}'^{\nu}}{\partial\text{x}^{\beta}}F^{\alpha\beta}\mathsf{e}_{\mu}\mathsf{e}_{\nu}.
\end{equation}
Employing Eq. (\ref{44}) twice, we convert the above relation to
\begin{equation}
F'=\tfrac{1}{2}(-\varepsilon)^{3}(\text{x}\mathsf{e}_{\alpha}\text{x})(\text{x}\mathsf{e}_{\beta}\text{x})F^{\alpha\beta}=-\varepsilon\text{x}^{2}\text{x}F\text{x}=-\varepsilon\Omega\text{x}F\text{x},\label{48}
\end{equation}
where 
\begin{equation}
\Omega:=\text{x}^{2}=\text{x}'^{-2}.
\end{equation}
Note that the results in Eqs. (\ref{46}) and (\ref{48}) are also
quite similar. 

Then, using the inverse of Eq. (\ref{14}), 
\begin{equation}
\text{x}=\varepsilon\text{x}'^{-2}\text{x}'\iff\text{x}^{\mu}=\varepsilon\text{x}'^{-2}\text{x}'^{\mu},
\end{equation}
we can re-express the results of Eqs. (\ref{46}) and (\ref{48})
in terms of the new coordinates, 
\begin{align}
 & A'=\text{x}'^{-4}\text{x}'A\text{x}'=\Omega^{2}\text{x}'A\text{x}',\label{51}\\
 & F'=-\varepsilon\text{x}'^{-6}\text{x}'F\text{x}'=-\varepsilon\Omega^{3}\text{x}'F\text{x}'.\label{52}
\end{align}

\subsection*{C. Formulas for SCT}

According to Eqs. (\ref{35}) and (\ref{46}), the transformation
of potential four-vector with respect to Eq. (\ref{18}) can be expressed
as 
\begin{align}
 & A'=\text{x}A\text{x},\nonumber \\
 & \tilde{A}=A',\label{53}\\
 & A''=\tilde{\text{x}}\tilde{A}\tilde{\text{x}}.\nonumber 
\end{align}
It is straightforward to write down the composition
\begin{equation}
A''=\tilde{\text{x}}\text{x}A\text{x}\tilde{\text{x}}=(1+\text{a}\text{x})A(1+\text{x}\text{a}),\label{54}
\end{equation}
where $\tilde{\text{x}}\text{x}=\varepsilon(1+\text{a}\text{x})\text{ and }\text{x}\tilde{\text{x}}=\varepsilon(1+\text{x}\text{a})$
have been used, and the arguments of $A''=A''(\text{x}'')$ and $A=A(\text{x})$
are related by Eq. (\ref{19}). 

In light of Eq. (\ref{51}), we can rewrite Eq. (\ref{54}) in terms
of the new coordinates,
\begin{equation}
A''=\text{x}'^{-4}\text{x}''^{-4}\text{x}''\text{x}'A\text{x}'\text{x}''=\Sigma{}^{2}(1-\text{x}''\text{a})A(1-\text{a}\text{x}''),\label{55}
\end{equation}
where $\Sigma$ is given in Eq. (\ref{33}).

Similarly, the SCT formulas of Faraday bivector can be derived from
Eqs. (\ref{42}), (\ref{48}), and (\ref{52}),
\begin{align}
F'' & =\text{x}^{2}\tilde{\text{x}}^{2}\tilde{\text{x}}\text{x}F\text{x}\tilde{\text{x}}=\Sigma(1+\text{a}\text{x})F(1+\text{x}\text{a})\label{56}\\
 & =\text{x}'^{-6}\text{x}''^{-6}\text{x}''\text{x}'F\text{x}'\text{x}''=\Sigma^{3}(1-\text{x}''\text{a})F(1-\text{a}\text{x}'').\label{57}
\end{align}

The Inversion and SCT formulas of current four-vector, which resemble
those of potential four-vector, are listed in Appendix II.

\subsection*{D. Review of Clifford algebra $C\!\ell_{3}$}

Serving as an alternative tool for electrodynamics \cite{key-6},
Clifford algebra $C\!\ell_{3}$ of Euclidean space is an eight-dimensional
vector space with the basis 
\begin{equation}
\{1;\hat{x},\hat{y},\hat{z};\hat{x}\hat{y},\hat{y}\hat{z},\hat{z}\hat{x};\hat{x}\hat{y}\hat{z}\},\label{58}
\end{equation}
where the geometric product is defined by
\begin{equation}
\hat{x}^{2}=\hat{y}^{2}=\hat{z}^{2}=1\text{ and }\hat{x}\hat{y}=-\hat{y}\hat{x},\text{ etc.}
\end{equation}
Hence for a Euclidean vector
\begin{equation}
\vec{U}^{2}=\vec{U\cdot}\vec{U}=U_{x}^{2}+U_{y}^{2}+U_{z}^{2}=:U^{2}.\label{60}
\end{equation}

In $C\!\ell_{3}$ formulation, the correspondents of position four-vector,
potential four-vector, and Faraday bivector of $C\!\ell_{1,3}$ formulation
are constructed as follows.
\begin{align}
 & \text{x}=t+\vec{r},\label{61}\\
 & A=A_{0}+\vec{A},\label{62}\\
 & F=E_{x}\hat{x}+E_{y}\hat{y}+E_{z}\hat{z}+B_{x}\hat{y}\hat{z}+B_{y}\hat{z}\hat{x}+B_{z}\hat{x}\hat{y}.\label{63}
\end{align}

Since the basis vector $\hat{x}\hat{y}\hat{z}$ satisfies
\begin{equation}
(\hat{x}\hat{y}\hat{z})^{2}=-1\text{ and (\ensuremath{\hat{x}\hat{y}\hat{z}})\ensuremath{\hat{x}}=\ensuremath{\hat{x}}(\ensuremath{\hat{x}\hat{y}\hat{z}})},\text{ etc.},
\end{equation}
it can be formally identified with $\sqrt{-1}=:i$. With this notation,
we rewrite Eqs. (\ref{58}) and (\ref{63}) respectively as
\begin{equation}
\{1;\hat{x},\hat{y},\hat{z};i\hat{z},i\hat{x},i\hat{y};i\},\label{65}
\end{equation}
\begin{equation}
\vec{F}=E_{x}\hat{x}+E_{y}\hat{y}+E_{z}\hat{z}+i(B_{x}\hat{x}+B_{y}\hat{y}+B_{z}\hat{z})=\vec{E}+i\vec{B}.\label{66}
\end{equation}
Although the symbols $\text{x},$ $A,$ and $F$ in Eqs. (\ref{61})$-$$(\ref{63})$
are the same as those in $C\!\ell_{1,3}$ formulation, we use $\vec{F}$
instead of $F$ in Eq. (\ref{66}) to emphasize it may be taken as
a complex-valued Euclidean vector, which will henceforward be called
Faraday vector.

Moreover, this $i$ notation enables us to express the geometric product
of two Euclidean vectors as
\begin{equation}
\vec{U}\vec{V}=\vec{U\cdot}\vec{V}+i\vec{U}\times\vec{V},\label{67}
\end{equation}
and it follows that
\begin{equation}
\vec{U}\vec{V}\vec{U}=2(\vec{U}\cdot\vec{V})\vec{U}-U^{2}\vec{V}.\label{68}
\end{equation}

\subsection*{E. Correspondence between $C\!\ell_{1,3}$ and $C\!\ell_{3}$ formulations}

In comparison, the covariant $C\!\ell_{1,3}$ formulation is suitable
for deriving elegant formulas, while the non-covariant $C\!\ell_{3}$
formulation is easier to manipulate in practical calculations. The
correspondence between the formulas of these two formulations are
listed here, the conversion rules are provided in Appendix I.

(i) Inversion:
\begin{align}
\text{Eqs. }(\ref{46})\text{ and }(\ref{51}) & \iff A'=\text{x}\bar{A}\text{x}=\omega^{2}\text{x}'\bar{A}\text{x}',\\
\text{Eqs. }(\ref{48})\text{ and }(\ref{52}) & \iff\vec{F}'=\varepsilon\omega\text{x}\vec{F}^{*}\bar{\text{x}}=\varepsilon\omega^{3}\text{x}'\vec{F}^{*}\bar{\text{x}}',\label{70}
\end{align}

\noindent where 
\begin{equation}
\omega:=\text{x}\bar{\text{x}}=\bar{\text{x}}\text{x}=(\text{x}'\bar{\text{x}}')^{-1}=(\bar{\text{x}}'\text{x}')^{-1},
\end{equation}
and the bar and asterisk symbols are defined via the following relations:
\begin{align}
 & \bar{\text{x}}=t-\vec{r},\nonumber \\
 & \bar{A}=A_{0}-\vec{A},\\
 & \vec{F}^{*}=\vec{E}-i\vec{B}.\nonumber 
\end{align}

(ii) SCT:
\begin{align}
\text{Eq. }(\ref{32}) & \iff\text{x}''=\sigma^{-1}(1+\text{a}\bar{\text{x}})\text{x}=\sigma^{-1}\text{x}(1+\bar{\text{x}}\text{a}),\\
\text{Eqs. }(\ref{54})\text{ and }(\ref{55}) & \iff A''=(1+\text{a}\bar{\text{x}})A(1+\bar{\text{x}}\text{a})=\sigma^{2}(1-\text{x}''\bar{\text{a}})A(1-\bar{\text{a}}\text{x}''),\\
\text{Eqs. }(\ref{56})\text{ and }(\ref{57}) & \iff\vec{F}''=\sigma(1+\text{a}\bar{\text{x}})\vec{F}(1+\text{x}\bar{\text{a}})=\sigma^{3}(1-\text{x}''\bar{\text{a}})\vec{F}(1-\text{a}\bar{\text{x}}''),\label{75}
\end{align}
where 
\begin{align}
\sigma & :=(1+\text{a}\bar{\text{x}})(1+\text{x}\bar{\text{a}})=(1+\bar{\text{a}}\text{x})(1+\bar{\text{x}}\text{a})\nonumber \\
 & =\big[(1-\text{x}''\bar{\text{a}})(1-\text{a}\bar{\text{x}}'')\big]^{-1}=\big[(1-\bar{\text{x}}''\text{a})(1-\bar{\text{a}}\text{x}'')\big]^{-1}.
\end{align}

Note that $\omega=\Omega$ and $\sigma=\Sigma$ when they are expressed
in terms of coordinates, i.e., 
\begin{equation}
\omega=t^{2}-r^{2}=(t'^{2}-r'^{2})^{-1},
\end{equation}
\begin{align}
\sigma & =1+2(\text{a}_{0}t-\vec{\text{a}}\cdot\vec{r})+(\text{a}_{0}^{2}-\vec{\text{a}}\cdot\vec{\text{a}})(t^{2}-r^{2})\nonumber \\
 & =\big[1-2(\text{a}_{0}t''-\vec{\text{a}}\cdot\vec{r\,}'')+(\text{a}_{0}^{2}-\vec{\text{a}}\cdot\vec{\text{a}})(t''^{2}-r''^{2})\big]^{-1}.
\end{align}

(iii) For completeness, we include the correspondence for parity transformation
formulas.
\begin{align}
\text{Eq. }(\ref{37}) & \iff\tilde{W}=\bar{W},\\
\text{Eq. }(\ref{43}) & \iff\vec{\tilde{F}}=-\vec{F}^{*}.
\end{align}

\section*{IV. APPLICATIONS OF THE FORMULAS}

\subsection*{A. Inversion and SCT of Lorentz invariants}

In electrodynamics, there exist two renowned Lorentz invariants 
\begin{equation}
I_{1}=\vec{E}\cdot\vec{E}-\vec{B}\cdot\vec{B}\text{ and }I_{2}=2\vec{E}\cdot\vec{B}
\end{equation}
that can be generated by Faraday vector via Eq. (\ref{60}), i.e.,
\begin{equation}
\vec{F}{}^{2}=\vec{F}\cdot\vec{F}=I_{1}+iI_{2}.\label{82}
\end{equation}

After Inversion, we can still define
\begin{equation}
I'_{1}+iI'_{2}:=\vec{F}'^{2}=\vec{E}'\cdot\vec{E}'-\vec{B}'\cdot\vec{B}'+2i\vec{E}'\cdot\vec{B}',\label{83}
\end{equation}
although $I'_{1}$ and $I'_{2}$ may not be invariant under Lorentz
transformation. On the other hand, we can use Eqs. (\ref{70}) and
(\ref{82}) to obtain
\begin{equation}
\vec{F}'{}^{2}=\omega^{2}\text{x}\vec{F}^{*}\bar{\text{x}}\text{x}\vec{F}^{*}\bar{\text{x}}=\omega^{4}\big(\vec{F}{}^{*}\big)^{2}=\omega^{4}(I_{1}-iI_{2}).\label{84}
\end{equation}
Comparing Eqs. (\ref{83}) and (\ref{84}), we find
\begin{equation}
I'_{1}=\omega^{4}I_{1}\text{ and }I'_{2}=-\omega^{4}I_{2}.\label{85}
\end{equation}

Similarly, the SCT formula Eq. (\ref{75}) leads to
\begin{equation}
\vec{F}''{}^{2}=\sigma^{2}(1+\text{a}\bar{\text{x}})\vec{F}(1+\text{x}\bar{\text{a}})(1+\text{a}\bar{\text{x}})\vec{F}(1+\text{x}\bar{\text{a}})=\sigma^{4}\vec{F}{}^{2},
\end{equation}
which yields
\begin{equation}
I''_{1}=\sigma^{4}I_{1}\text{ and }I''_{2}=\sigma^{4}I_{2}.\label{87}
\end{equation}

As a comparison, we sketch the derivations of $I'_{1}$ and $I'_{2}$
in tensorial formulation \cite{key-11}: 
\begin{equation}
I_{1}=-\tfrac{1}{2}F^{\alpha\beta}F_{\alpha\beta}\Longrightarrow I'_{1}=-\tfrac{1}{2}F'^{\mu\nu}F'_{\mu\nu};\label{88}
\end{equation}
\begin{equation}
I_{2}=-\tfrac{1}{4}\epsilon^{\alpha\beta\gamma\delta}F_{\alpha\beta}F_{\gamma\delta}\Longrightarrow I'_{2}=-\tfrac{1}{4}\epsilon'^{\mu\nu\rho\sigma}F'_{\mu\nu}F'_{\rho\sigma},\label{89}
\end{equation}
where $\epsilon^{\alpha\beta\gamma\delta}$ is the Levi-Civita symbol
that transforms as
\begin{equation}
\epsilon'^{\mu\nu\rho\sigma}=\det\bigg[\frac{\partial\text{x}}{\partial\text{x}'}\bigg]\frac{\partial\text{x}'^{\mu}}{\partial\text{x}^{\alpha}}\frac{\partial\text{x}'^{\nu}}{\partial\text{x}^{\beta}}\frac{\partial\text{x}'^{\rho}}{\partial\text{x}^{\gamma}}\frac{\partial\text{x}'^{\sigma}}{\partial\text{x}^{\delta}}\epsilon^{\alpha\beta\gamma\delta}.
\end{equation}
Invoking Eqs. (\ref{25}) and (\ref{26}), it is easy to obtain
\begin{equation}
I'_{1}=\Lambda^{4}I_{1}=\omega^{4}I_{1}
\end{equation}
from Eq. (\ref{88}). Nevertheless, Eq. (\ref{89}) leads to
\begin{equation}
I'_{2}=\det\bigg[\frac{\partial\text{x}}{\partial\text{x}'}\bigg]I_{2},\label{92}
\end{equation}
and it is laborious to show the Jacobian determinant equals $-\omega^{4}.$

\subsection*{B. Inversion of an electromagnetic field}

In this example, we demonstrate how to expand $\vec{F}'$ in Eq. (\ref{70})
to obtain the expression for Inversion of a given electromagnetic
field.

To express the result in terms of the original coordinates, we use
the formula
\begin{align}
\vec{F}' & =\varepsilon\omega\text{x}\vec{F}^{*}\bar{\text{x}}=\varepsilon\omega(t+\vec{r}\,)\vec{F}^{*}(t-\vec{r}\,)\label{93}\\
 & =\varepsilon\omega[(t^{2}+r^{2})\vec{F}^{*}-2(\vec{r}\cdot\vec{F}^{*})\vec{r}+2it(\vec{r}\times\vec{F}^{*})],\label{94}
\end{align}
where Eqs. (\ref{67}) and (\ref{68}) have been used in the derivation
of Eq. (\ref{94}). Alternatively, we can express the result in terms
of the new coordinates, 
\begin{align}
\vec{F}' & =\varepsilon\omega^{3}\text{x}'\vec{F}^{*}\bar{\text{x}}'=\varepsilon\omega^{3}(t'+\vec{r}\,')\vec{F}^{*}(t'-\vec{r}\,')\label{95}\\
 & =\varepsilon\omega^{3}[(t'^{2}+r'^{2})\vec{F}^{*}-2(\vec{r}\,'\cdot\vec{F}^{*})\vec{r}\,'+2it'(\vec{r}\,'\times\vec{F}^{*})].\label{96}
\end{align}
 Note that Eqs. (\ref{94}) and (\ref{96}) are isomorphic except
for the powers of $\omega$.

It is straightforward to separate Eq. (\ref{94}) or Eq. (\ref{96})
into electric and magnetic parts. In order to compare with existing
studies \cite{key-1,key-12}, we list the result of Eq. (\ref{94})
as follows.
\begin{align}
 & \vec{E}'=\varepsilon(t^{2}-r^{2})[(t^{2}+r^{2})\vec{E}-2(\vec{r}\cdot\vec{E})\vec{r}+2t(\vec{r}\times\vec{B})];\label{97}\\
 & \vec{B}'=\varepsilon(t^{2}-r^{2})[-(t^{2}+r^{2})\vec{B}+2(\vec{r}\cdot\vec{B})\vec{r}+2t(\vec{r}\times\vec{E})],\label{98}
\end{align}
or, equivalently,
\begin{align}
 & \vec{E}'=\varepsilon(t^{2}-r^{2})[(t^{2}-r^{2})\vec{E}-2\vec{r}\times(\vec{r}\times\vec{E})+2t(\vec{r}\times\vec{B})];\label{99}\\
 & \vec{B}'=\varepsilon(t^{2}-r^{2})[-(t^{2}-r^{2})\vec{B}+2\vec{r}\times(\vec{r}\times\vec{B})+2t(\vec{r}\times\vec{E})].\label{100}
\end{align}
It is apparent that, since $\vec{F}^{*}=\vec{E}-i\vec{B}$ and $i\vec{F}^{*}=\vec{B}+i\vec{E}$,
the inverse duality transformation $(\vec{E},\vec{B})\longrightarrow(-\vec{B},\vec{E})$
leads to Eq. (\ref{97})$\longrightarrow$ Eq. (\ref{98}) and Eq.
(\ref{99})$\longrightarrow$ Eq. (\ref{100}).

The tensorial counterpart of Eq. (\ref{94}) takes the form \cite{key-8}
\begin{equation}
F'^{\mu\nu}=-\varepsilon\big[\text{x}^{4}F^{\mu\nu}+2\text{x}^{2}(\text{x}^{\mu}F^{\nu\rho}-\text{x}^{\nu}F^{\mu\rho})\text{x}_{\rho}\big],\label{101}
\end{equation}
which can be obtained by substituting Eqs. (\ref{16}), (\ref{17}),
and (\ref{28}) into Eq. (\ref{25}). Although Eq. (\ref{101}) is
equivalent to Eq. (\ref{94}), there is no simple way to separate
it into electric and magnetic parts, nor can we see the electro-magnetic
duality directly from this expression. 

Note that if we employ the covariant formula in Eq. (\ref{48}) instead
of the non-covariant one in Eq. (\ref{70}), the expansion result
will be essentially Eq. (\ref{101}), i.e., $F'=\tfrac{1}{2}F'^{\mu\nu}\mathsf{e}_{\mu}\mathsf{e}_{\nu}$.

\subsection*{C. SCT of an electromagnetic field}

The goal of this example is to expand $\vec{F}''=\vec{E}''+i\vec{B}''$
in terms of the original coordinates, and the strategy is to imitate
the previous example as much as possible. Hence we start with expressing
$\vec{F}''$ in Eq. (\ref{75}) as 
\begin{equation}
\vec{F}''=\sigma(u+\vec{v})\vec{F}(u-\vec{v}),\label{102}
\end{equation}
where
\begin{equation}
u=1+\tfrac{1}{2}(\text{a}\bar{\text{x}}+\text{x}\bar{\text{a}})=1+\text{a}_{0}t-\vec{\text{a}}\cdot\vec{r}
\end{equation}
is a scalar, and
\begin{equation}
\vec{v}=\tfrac{1}{2}(\text{a}\bar{\text{x}}-\text{x}\bar{\text{a}})=t\vec{\text{a}}-\text{a}_{0}\vec{r}-i\vec{\text{a}}\times\vec{r}
\end{equation}
is a complex-valued Euclidean vector. 

From Eqs. (\ref{93}) and (\ref{94}), the expansion of Eq. (\ref{102})
takes the form
\begin{equation}
\vec{F}''=\sigma[(u^{2}+v^{2})\vec{F}-2(\vec{v}\cdot\vec{F})\vec{v}+2iu(\vec{v}\times\vec{F})],\label{105}
\end{equation}
where
\begin{equation}
v^{2}=(t\vec{\text{a}}-\text{a}_{0}\vec{r})^{2}-(\vec{\text{a}}\times\vec{r})^{2}=(t\vec{\text{a}}-\text{a}_{0}\vec{r})\cdot(t\vec{\text{a}}-\text{a}_{0}\vec{r})-(\vec{\text{a}}\times\vec{r})\cdot(\vec{\text{a}}\times\vec{r}).
\end{equation}

For the sake of clarity, we separate Eq. (\ref{105}) into three parts,
\begin{equation}
\vec{F}''=\vec{F}''_{1}+\vec{F}''_{2}+\vec{F}''_{3},
\end{equation}
where
\begin{equation}
\vec{F}''_{1}=\vec{E}''_{1}+i\vec{B}''_{1}=\sigma(u^{2}+v^{2})\vec{F},\label{108}
\end{equation}
\begin{equation}
\vec{F}''_{2}=\vec{E}''_{2}+i\vec{B}''_{2}=-2\sigma(\vec{v}\cdot\vec{F})\vec{v},\label{109}
\end{equation}
\begin{equation}
\vec{F}''_{3}=\vec{E}''_{3}+i\vec{B}''_{3}=2i\sigma u(\vec{v}\times\vec{F}).\label{110}
\end{equation}

For $\vec{F}''_{1}$ in Eq. (\ref{108}), we can directly write down
the result
\begin{equation}
\vec{E}''_{1}=\sigma\big[(1+\text{a}_{0}t-\vec{\text{a}}\cdot\vec{r})^{2}+(t\vec{\text{a}}-\text{a}_{0}\vec{r})^{2}-(\vec{\text{a}}\times\vec{r})^{2}\big]\vec{E};\label{111}
\end{equation}
\begin{equation}
\vec{B}''_{1}=\sigma\big[(1+\text{a}_{0}t-\vec{\text{a}}\cdot\vec{r})^{2}+(t\vec{\text{a}}-\text{a}_{0}\vec{r})^{2}-(\vec{\text{a}}\times\vec{r})^{2}\big]\vec{B}.\label{112}
\end{equation}

For $\vec{F}''_{2}$ in Eq. (\ref{109}), we first calculate
\begin{align}
(\vec{v}\cdot\vec{F})\vec{v} & =\big\{ t(\vec{\text{a}}\cdot\vec{E})-\text{a}_{0}(\vec{r}\cdot\vec{E})+(\vec{\text{a}}\times\vec{r})\cdot\vec{B}\nonumber \\
 & +i[t(\vec{\text{a}}\cdot\vec{B})-\text{a}_{0}(\vec{r}\cdot\vec{B})-(\vec{\text{a}}\times\vec{r})\cdot\vec{E}]\big\}[t\vec{\text{a}}-\text{a}_{0}\vec{r}-i\vec{\text{a}}\times\vec{r}\,],
\end{align}
it then follows that
\begin{align}
\vec{E}''_{2} & =-2\sigma[t(\vec{\text{a}}\cdot\vec{E})-\text{a}_{0}(\vec{r}\cdot\vec{E})+(\vec{\text{a}}\times\vec{r})\cdot\vec{B}](t\vec{\text{a}}-\text{a}_{0}\vec{r})\nonumber \\
 & -2\sigma[t(\vec{\text{a}}\cdot\vec{B})-\text{a}_{0}(\vec{r}\cdot\vec{B})-(\vec{\text{a}}\times\vec{r})\cdot\vec{E}]\vec{\text{a}}\times\vec{r};\label{114}\\
\vec{B}''_{2} & =-2\sigma[t(\vec{\text{a}}\cdot\vec{B})-\text{a}_{0}(\vec{r}\cdot\vec{B})-(\vec{\text{a}}\times\vec{r})\cdot\vec{E}](t\vec{\text{a}}-\text{a}_{0}\vec{r})\nonumber \\
 & +2\sigma[t(\vec{\text{a}}\cdot\vec{E})-\text{a}_{0}(\vec{r}\cdot\vec{E})+(\vec{\text{a}}\times\vec{r})\cdot\vec{B}]\vec{\text{a}}\times\vec{r}.\label{115}
\end{align}

Finally for $\vec{F}''_{3}$ in Eq. (\ref{110}), by applying the
formula
\begin{equation}
(\vec{\text{a}}\times\vec{r})\times\vec{U}=(\vec{\text{a}}\cdot\vec{U})\vec{r}-(\vec{r}\cdot\vec{U})\vec{\text{a}}
\end{equation}
to
\begin{equation}
\vec{v}\times\vec{F}=[(t\vec{\text{a}}-\text{a}_{0}\vec{r})\times\vec{E}+(\vec{\text{a}}\times\vec{r})\times\vec{B}]+i[(t\vec{\text{a}}-\text{a}_{0}\vec{r})\times\vec{B}-(\vec{\text{a}}\times\vec{r})\times\vec{E}],
\end{equation}
we obtain
\begin{equation}
\vec{E}''_{3}=2\sigma(1+\text{a}_{0}t-\vec{\text{a}}\cdot\vec{r})[(\vec{\text{a}}\cdot\vec{E})\vec{r}-(\vec{r}\cdot\vec{E})\vec{\text{a}}-t\vec{\text{a}}\times\vec{B}+\text{a}_{0}\vec{r}\times\vec{B}];\label{118}
\end{equation}
\begin{equation}
\vec{B}''_{3}=2\sigma(1+\text{a}_{0}t-\vec{\text{a}}\cdot\vec{r})[(\vec{\text{a}}\cdot\vec{B})\vec{r}-(\vec{r}\cdot\vec{B})\vec{\text{a}}+t\vec{\text{a}}\times\vec{E}-\text{a}_{0}\vec{r}\times\vec{E}].\label{119}
\end{equation}

In summary, the expressions for the transformed electric and magnetic
fields are respectively \cite{key-13,key-14} 
\begin{equation}
\vec{E}''=\vec{E}''_{1}+\vec{E}''_{2}+\vec{E}''_{3};\label{120}
\end{equation}
\begin{equation}
\vec{B}''=\vec{B}''_{1}+\vec{B}''_{2}+\vec{B}''_{3}.\label{121}
\end{equation}
Since $\vec{F}=\vec{E}+i\vec{B}$ and $i\vec{F}=-\vec{B}+i\vec{E}$,
the duality transformation $(\vec{E},\vec{B})\longrightarrow(\vec{B},-\vec{E})$
leads to Eq. (\ref{120})$\longrightarrow$ Eq. (\ref{121}).

Comparing the two expressions in Eq. (\ref{75}), we find that to
express $\vec{E}''$ and $\vec{B}''$ in terms of the new coordinates,
one only has to replace each explicit $t$ in Eqs. (\ref{111}), (\ref{112}),
(\ref{114}), (\ref{115}), (\ref{118}) and (\ref{119}) by $t''$,
and each explicit $\vec{r}$ therein by $\vec{r}\,''$, then make
the following changes:
\begin{align}
\forall\ \sigma & \longrightarrow\sigma^{3},\\
\forall\ 1+\text{a}_{0}t''-\vec{\text{a}}\cdot\vec{r\,}'' & \longrightarrow1-\text{a}_{0}t''+\vec{\text{a}}\cdot\vec{r\,}''.
\end{align}

In tensorial formulation as well as $C\!\ell_{1,3}$ formulation,
the corresponding calculations are much lengthier, and the result
\cite{key-7}
\begin{align}
F''^{\mu\nu} & =\sigma^{2}F^{\mu\nu}-2\sigma\big(\text{a}^{\mu}F^{\nu\rho}-\text{a}^{\nu}F^{\mu\rho}\big)[\text{x}_{\rho}+2(\text{a}\cdot\text{x})\text{x}_{\rho}-\text{x}^{2}\text{a}_{\rho}]\nonumber \\
 & +2\sigma\big(\text{x}^{\mu}F^{\nu\rho}-\text{x}^{\nu}F^{\mu\rho}\big)(\text{a}^{2}\text{x}_{\rho}+\text{a}_{\rho})+4\sigma\big(\text{a}^{\mu}\text{x}^{\nu}-\text{a}^{\nu}\text{x}^{\mu}\big)\text{a}_{\rho}F^{\rho\sigma}\text{x}_{\sigma}\label{124}
\end{align}
is not easy to be separated into electric and magnetic parts. Moreover,
even if we accomplish the separation \cite{key-3}, a lot of further
work is needed to rearrange the result into the forms of Eqs. (\ref{111}),
(\ref{112}), (\ref{114}), (\ref{115}), (\ref{118}) and (\ref{119}).

\section*{V. DISCUSSION AND CONCLUSION}

In this research we derived a set of Clifford-algebraic formulas for
Inversion and SCT of the quantities related to Maxwell's equations.
The advantages of these formulas are threefold: 

(i) As evidenced by the three illustrative examples in Sec. IV, these
formulas enjoy superiority in conciseness and convenience compared
to the vector-tensorial formulas. This strengthens our confidence
in our results in the face of possible discrepancies with existing
studies \cite{key-1,key-12,key-13,key-14}. 

(ii) These formulas neatly integrate the covariant and contravariant
formulas in vector-tensorial formulation. For example, owing to $A'^{\mu}\mathsf{e}_{\mu}=A'_{\mu}\mathsf{e}^{\mu}$,
the $C\!\ell_{1,3}$ formula $A'=\text{x}A\text{x}$ is mathematically
equivalent to both 
\begin{equation}
A'^{\mu}=\vartheta\text{x}^{4}\frac{\partial\text{x}'^{\mu}}{\partial\text{x}^{\alpha}}A^{\alpha}=-\text{x}^{2}A^{\mu}+2(\text{x}\cdot A)\text{x}^{\mu}
\end{equation}
and 
\begin{equation}
A'_{\mu}=\eta_{\mu\nu}A'^{\nu}=\vartheta\frac{\partial\text{x}^{\alpha}}{\partial\text{x}'^{\mu}}A_{\alpha}=-\text{x}^{2}A_{\mu}+2(\text{x}\cdot A)\text{x}_{\mu}.
\end{equation}
Similarly, $A''=(1+\text{a}\text{x})A(1+\text{x}\text{a})$ is equivalent
to 
\begin{align}
A''^{\mu} & =\big[\sigma\delta_{\alpha}^{\mu}-2(\text{a}_{\alpha}\text{x}^{\mu}-\text{x}_{\alpha}\text{a}^{\mu})+4(\text{a}\cdot\text{x})\text{x}_{\alpha}\text{a}^{\mu}-2(\text{a}^{2}\text{x}_{\alpha}\text{x}^{\mu}+\text{x}^{2}\text{a}_{\alpha}\text{a}^{\mu})\big]A^{\alpha}\nonumber \\
 & =\sigma A^{\mu}-2\big[\text{a}\cdot A+\text{a}^{2}(\text{x}\cdot A)\big]\text{x}^{\mu}+2\big[\text{x}\cdot A-\text{x}^{2}(\text{a}\cdot A)+2(\text{a}\cdot\text{x})(\text{x}\cdot A)\big]\text{a}^{\mu}
\end{align}
and its covariant counterpart $A''_{\mu}=\eta_{\mu\nu}A''^{\nu}$. 

For the same reason $F'=-\varepsilon\Omega\text{x}F\text{x}$ is equivalent
to $F'^{\mu\nu}$ in Eq. (\ref{101}) and $F'_{\mu\nu}=\eta_{\mu\rho}\eta_{\nu\sigma}F'^{\rho\sigma}$,
while $F''=\Sigma(1+\text{a}\text{x})F(1+\text{x}\text{a})$ to $F''^{\mu\nu}$
in Eq. (\ref{124}) and its covariant counterpart.

(iii) In the literature, there is no lack of introducing higher-dimensional
spacetime to realize conformal transformation, either in vector-tensorial
formulation \cite{key-3,key-7,key-11} or Clifford-algebraic formulation
\cite{key-15}. In contrast, our formulas are constructed and operated
entirely in four dimensions. Therefore, as listed in Appendix II,
all of the fundamental conformal transformations of the quantities
related to Maxwell's equations can be expressed as Clifford-algebraic
formulas in ordinary spacetime.

In conclusion, the formulas we derived offer a streamlined mathematical
framework for the major nonlinear conformal transformations in electrodynamics.
This not only paves the way for future research, but also holds the
potential for bringing new insight into existing studies.

\section*{Appendix I: $C\!\ell_{1,3}\protect\longrightarrow C\!\ell_{3}$ CONVERSION}

\noindent (i) Basic rules:
\[
\mathsf{e}_{1}\mathsf{e}_{0}\longrightarrow\hat{x},\ \mathsf{e}_{2}\mathsf{e}_{0}\longrightarrow\hat{y},\ \mathsf{e}_{3}\mathsf{e}_{0}\longrightarrow\hat{z},\text{ and }\mathsf{e}_{0}^{2}=1.
\]
(ii) Four-vectors: 
\begin{align*}
\text{x}\mathsf{e}_{0} & \longrightarrow\text{x}=t+\vec{r},\\
\mathsf{\mathsf{e}_{0}\text{x}} & \longrightarrow\bar{\text{x}}=t-\vec{r},\\
A\mathsf{e}_{0} & \longrightarrow A=A_{0}+\vec{A},\\
\mathsf{e}_{0}A & \longrightarrow\bar{A}=A_{0}-\vec{A}.
\end{align*}
(iii) Faraday (bi-)vector:
\begin{align*}
F & =E_{x}\mathsf{e}_{1}\mathsf{e}_{0}+E_{y}\mathsf{e}_{2}\mathsf{e}_{0}+E_{z}\mathsf{e}_{3}\mathsf{e}_{0}+B_{x}\mathsf{e}_{3}\mathsf{e}_{2}+B_{y}\mathsf{e}_{1}\mathsf{e}_{3}+B_{z}\mathsf{e}_{2}\mathsf{e}_{1}\\
 & =E_{x}\mathsf{e}_{1}\mathsf{e}_{0}+E_{y}\mathsf{e}_{2}\mathsf{e}_{0}+E_{z}\mathsf{e}_{3}\mathsf{e}_{0}+B_{x}\mathsf{e}_{3}\mathsf{e}_{0}\mathsf{e}_{0}\mathsf{e}_{2}+B_{y}\mathsf{e}_{1}\mathsf{e}_{0}\mathsf{e}_{0}\mathsf{e}_{3}+B_{z}\mathsf{e}_{2}\mathsf{e}_{0}\mathsf{e}_{0}\mathsf{e}_{1}\\
 & \longrightarrow E_{x}\hat{x}+E_{y}\hat{y}+E_{z}\hat{z}+B_{x}\hat{y}\hat{z}+B_{y}\hat{z}\hat{x}+B_{z}\hat{x}\hat{y}=\vec{E}+i\vec{B}=\vec{F}.
\end{align*}
(iv) Geometric products of two four-vectors:
\[
\text{x}\text{y}=\text{x}\mathsf{e}_{0}\mathsf{e}_{0}\text{y}\longrightarrow\text{x}\bar{\text{y}},
\]
\[
\text{x}\cdot\text{x}=\text{x}^{2}=\text{x}\mathsf{e}_{0}\mathsf{e}_{0}\text{x}=\mathsf{e}_{0}\text{x}\text{x}\mathsf{e}_{0}\longrightarrow\text{x}\bar{\text{x}}=\bar{\text{x}}\text{x},
\]
\[
2\text{x}\cdot\text{y}=\text{x}\text{y}+\text{y}\text{x}\longrightarrow\text{x}\bar{\text{y}}+\text{y}\bar{\text{x}}=\bar{\text{x}}\text{y}+\bar{\text{y}}\text{x}.
\]
(v) Faraday (bi-)vector sandwiched by two four-vectors:
\[
\text{x}F\text{y}=\text{x}\mathsf{e}_{0}(\mathsf{e}_{0}F\mathsf{e}_{0})\mathsf{e}_{0}\text{y}\longrightarrow\text{x}(-\vec{E}+i\vec{B})\bar{\text{y}}=-\text{x}\vec{F}^{*}\bar{\text{y}}.
\]

\section*{Appendix II: SUMMARY OF FORMULAS }

\subsection*{A. $C\!\ell_{1,3}$ formulation:}

\begin{align*}
\text{Position four-vector: } & \text{x}=\text{x}^{\alpha}\mathsf{e}_{\alpha},\\
\text{Potential four-vector: } & A(\text{x})=A^{\alpha}\mathsf{e}_{\alpha},\\
\text{Current four-vector: } & J(\text{x})=J^{\alpha}\mathsf{e}_{\alpha},\\
\text{Faraday bivector: } & F(\text{x})=\tfrac{1}{2}F^{\alpha\beta}\mathsf{e}_{\alpha}\mathsf{e}_{\beta}.
\end{align*}
($\bm{\mathit{1}}$) Four-dimensional dilation: 
\[
\tilde{\text{x}}=\lambda^{-1}\text{x},\ \tilde{A}=\lambda A,\ \tilde{J}=\lambda^{3}J,\ \tilde{F}=\lambda^{2}F.
\]

\noindent ($\bm{\mathit{2}}$) Four-dimensional translation: 

\[
\tilde{\text{x}}=\text{x}+\text{b},\ \tilde{A}=A,\ \tilde{J}=J,\ \tilde{F}=F.
\]
\\
\\

\noindent ($\bm{\mathit{3}}$) Lorentz transformation$=$Four-dimensional
orthogonal transformation: 
\[
L:=\exp(\alpha_{x}\mathsf{e}_{1}\mathsf{e}_{0}+\alpha_{y}\mathsf{e}_{2}\mathsf{e}_{0}+\alpha_{z}\mathsf{e}_{3}\mathsf{e}_{0}+\theta_{x}\mathsf{e}_{3}\mathsf{e}_{2}+\theta_{y}\mathsf{e}_{1}\mathsf{e}_{3}+\theta_{z}\mathsf{e}_{2}\mathsf{e}_{1}).
\]

\noindent ($\bm{\mathit{3}.1}$) Proper orthochronous $(\vartheta=1)$:
\[
\tilde{\text{x}}=L\text{x}L^{-1},\ \tilde{A}=LAL^{-1},\ \tilde{J}=LJL^{-1},\ \tilde{F}=LFL^{-1}.
\]

\noindent ($\bm{\mathit{3}.2}$) Improper orthochronous $(\vartheta=1)$:
\[
\tilde{\text{x}}=\mathsf{e}_{0}L\text{x}L^{-1}\mathsf{e}_{0},\ \tilde{A}=\mathsf{e}_{0}LAL^{-1}\mathsf{e}_{0},\ \tilde{J}=\mathsf{e}_{0}LJL^{-1}\mathsf{e}_{0},\ \tilde{F}=\mathsf{e}_{0}LFL^{-1}\mathsf{e}_{0}.
\]

\noindent ($\bm{\mathit{3}.3}$) Improper antichronous $(\vartheta=-1)$:
\[
\tilde{\text{x}}=-\mathsf{e}_{0}L\text{x}L^{-1}\mathsf{e}_{0},\ \tilde{A}=\mathsf{e}_{0}LAL^{-1}\mathsf{e}_{0},\ \tilde{J}=\mathsf{e}_{0}LJL^{-1}\mathsf{e}_{0},\ \tilde{F}=-\mathsf{e}_{0}LFL^{-1}\mathsf{e}_{0}.
\]

\noindent ($\bm{\mathit{3}.4}$) Proper antichronous $(\vartheta=-1)$:
\[
\tilde{\text{x}}=-L\text{x}L^{-1},\ \tilde{A}=LAL^{-1},\ \tilde{J}=LJL^{-1},\ \tilde{F}=-LFL^{-1}.
\]

\noindent ($\bm{\mathit{4}}'$) Conformal inversion: 
\begin{align*}
\Omega & =\text{x}^{2}=\text{x}'^{-2};\\
\text{x}' & =\varepsilon\Omega^{-1}\text{x},\\
A' & =\text{x}A\text{x}=\Omega^{2}\text{x}'A\text{x}',\\
J' & =\Omega^{2}\text{x}J\text{x}=\Omega^{4}\text{x}'J\text{x}',\\
F' & =-\varepsilon\Omega\text{x}F\text{x}=-\varepsilon\Omega^{3}\text{x}'F\text{x}'.
\end{align*}
($\bm{\mathit{4}}$) Special conformal transformation (SCT): 
\begin{align*}
\Sigma & =(1+\text{a}\text{x})(1+\text{x}\text{a})=\big[(1-\text{x}''\text{a})(1-\text{a}\text{x}'')\big]^{-1};\\
\text{x}'' & =\Sigma^{-1}(1+\text{a}\text{x})\text{x}=\Sigma^{-1}\text{x}(1+\text{x}\text{a}),\\
A'' & =(1+\text{a}\text{x})A(1+\text{x}\text{a})=\Sigma^{2}(1-\text{x}''\text{a})A(1-\text{a}\text{x}''),\\
J'' & =\Sigma^{2}(1+\text{a}\text{x})J(1+\text{x}\text{a})=\Sigma^{4}(1-\text{x}''\text{a})J(1-\text{a}\text{x}''),\\
F'' & =\Sigma(1+\text{a}\text{x})F(1+\text{x}\text{a})=\Sigma^{3}(1-\text{x}''\text{a})F(1-\text{a}\text{x}'').
\end{align*}

\subsection*{B. $C\!\ell_{3}$ formulation:}

\begin{align*}
\text{Position four-vector: } & \text{x}=t+\vec{r},\\
\text{Potential four-vector: } & A(\text{x})=A_{0}+\vec{A},\\
\text{Current four-vector: } & J(\text{x})=J_{0}+\vec{J},\\
\text{Faraday vector: } & \vec{F}(\text{x})=\vec{E}+i\vec{B}.
\end{align*}
($\bm{\mathit{1}}$) Four-dimensional dilation: $\tilde{\text{x}}=\lambda^{-1}\text{x},\ \tilde{A}=\lambda A,\ \tilde{J}=\lambda^{3}J,\ \vec{\tilde{F}}=\lambda^{2}\vec{F}.$\\

\noindent ($\bm{\mathit{2}}$) Four-dimensional translation: $\tilde{\text{x}}=\text{x}+\text{b},\ \tilde{A}=A,\ \tilde{J}=J,\ \vec{\tilde{F}}=\vec{F}.$\\

\noindent ($\bm{\mathit{3}}$) Lorentz transformation: $L=\exp\big(\vec{\alpha}+i\vec{\theta}\,\big)$.

\noindent ($\bm{\mathit{3}.1}$) Proper orthochronous: $\tilde{\text{x}}=L\text{x}L^{*},\ \tilde{A}=LAL^{*},\ \tilde{J}=LJL^{*},\ \vec{\tilde{F}}=L\vec{F}\bar{L}.$ 

\noindent ($\bm{\mathit{3}.2}$) Improper orthochronous: $\tilde{\text{x}}=\bar{L}^{*}\bar{\text{x}}\bar{L},\ \tilde{\text{A}}=\bar{L}^{*}\bar{\text{A}}\bar{L},\ \tilde{\text{J}}=\bar{L}^{*}\bar{\text{J}}\bar{L},\ \vec{\tilde{F}}=-\bar{L}^{*}\vec{F}^{*}L^{*}.$

\noindent ($\bm{\mathit{3}.3}$) Improper antichronous: $\tilde{\text{x}}=-\bar{L}^{*}\bar{\text{x}}\bar{L},\ \tilde{\text{A}}=\bar{L}^{*}\bar{\text{A}}\bar{L},\ \tilde{\text{J}}=\bar{L}^{*}\bar{\text{J}}\bar{L},\ \vec{\tilde{F}}=\bar{L}^{*}\vec{F}^{*}L^{*}$.

\noindent ($\bm{\mathit{3}.4}$) Proper antichronous: $\tilde{\text{x}}=-L\text{x}L^{*},\ \tilde{A}=LAL^{*},\ \tilde{J}=LJL^{*},\ \vec{\tilde{F}}=-L\vec{F}\bar{L}$.
\\

\noindent ($\bm{\mathit{4}}'$) Conformal inversion: 
\begin{align*}
\omega & =\text{x}\bar{\text{x}}=\bar{\text{x}}\text{x}=(\text{x}'\bar{\text{x}}')^{-1}=(\bar{\text{x}}'\text{x}')^{-1};\\
\text{x}' & =\varepsilon\omega^{-1}\text{x},\\
A' & =\text{x}\bar{A}\text{x}=\omega^{2}\text{x}'\bar{A}\text{x}',\\
J' & =\omega^{2}\text{x}\bar{J}\text{x}=\omega^{4}\text{x}'\bar{J}\text{x}',\\
\vec{F}' & =\varepsilon\omega\text{x}\vec{F}^{*}\bar{\text{x}}=\varepsilon\omega^{3}\text{x}'\vec{F}^{*}\bar{\text{x}}'.
\end{align*}
($\bm{\mathit{4}}$) Special conformal transformation (SCT): 
\begin{align*}
\sigma & =(1+\text{a}\bar{\text{x}})(1+\text{x}\bar{\text{a}})=(1+\bar{\text{a}}\text{x})(1+\bar{\text{x}}\text{a})\\
 & =\big[(1-\text{x}''\bar{\text{a}})(1-\text{a}\bar{\text{x}}'')\big]^{-1}=\big[(1-\bar{\text{x}}''\text{a})(1-\bar{\text{a}}\text{x}'')\big]^{-1};\\
\text{x}'' & =\sigma^{-1}(1+\text{a}\bar{\text{x}})\text{x}=\sigma^{-1}\text{x}(1+\bar{\text{x}}\text{a}),\\
A'' & =(1+\text{a}\bar{\text{x}})A(1+\bar{\text{x}}\text{a})=\sigma^{2}(1-\text{x}''\bar{\text{a}})A(1-\bar{\text{a}}\text{x}''),\\
J'' & =\sigma^{2}(1+\text{a}\bar{\text{x}})J(1+\bar{\text{x}}\text{a})=\sigma^{4}(1-\text{x}''\bar{\text{a}})J(1-\bar{\text{a}}\text{x}''),\\
\vec{F}'' & =\sigma(1+\text{a}\bar{\text{x}})\vec{F}(1+\text{x}\bar{\text{a}})=\sigma^{3}(1-\text{x}''\bar{\text{a}})\vec{F}(1-\text{a}\bar{\text{x}}'').
\end{align*}

In both formulations, the similarity between the expressions of $A'$
and $J'$ can be deduced by comparing Eq. (\ref{21}) with Eq. (\ref{23}),
or Eq. (\ref{22}) with Eq. (\ref{24}). The same argument applies
to the similarity between the expressions of $A''$ and $J''$.


\begin{thebibliography}{10}
\bibitem{key-1}E. Cunningham, \textquotedblleft The principle of
relativity in electrodynamics and an extension thereof,\textquotedblright{}
Proc. London Math. Soc. \textbf{8}, 77-98 (1910). 

\bibitem{key-2}H. Bateman, \textquotedblleft The transformation of
the electrodynamical equations,\textquotedblright{} Proc. London Math.
Soc. \textbf{8}, 223-264 (1910).

\bibitem{key-3}C. Codirla and H. Osborn, \textquotedblleft Conformal
invariance and electrodynamics: applications and general formalism,\textquotedblright{}
Ann. Phys. (N.Y.) \textbf{260}, 91-116 (1997), and references therein.

\bibitem{key-4}C. Doran and A. Lasenby, \textit{Geometric Algebra
for Physicists} (Cambridge University Press, 2003).

\bibitem{key-5}D. Hestenes, \textit{Space-Time Algebra}, 2nd ed.
(Birkha\"user, 2015).

\bibitem{key-6}W. E. Baylis, \textit{Electrodynamics: A Modern Geometric
Approach} (Birkha\"user, 1998).

\bibitem{key-7}A. O. Barut and R. B. Haugen, \textquotedblleft Theory
of the conformally invariant mass,\textquotedblright{} Ann. Phys.
(N.Y.) \textbf{71}, 519-541 (1972).

\bibitem{key-8}L. Yeh, \textquotedblleft Conformal transformation
and Maxwell's equations,\textquotedblright{} hal-04281513 (2023).
Note that some notations therein are different from those in this
paper.

\bibitem{key-9}M. Schottenloher, \textit{A Mathematical Introduction
to Conformal Field Theory}, 2nd ed. (Springer, 2008), p. 18.

\bibitem{key-10}H. Laue, \textquotedblleft Causality and the spontaneous
break-down of conformal symmetry,\textquotedblright{} Nuovo \\
Cimento \textbf{10B}, 283-290 (1972).

\bibitem{key-11}S. Duplij, G. A. Goldin, and V. Shtelen, \textquotedblleft Conformal
inversion and Maxwell field invariants in four- and six-dimensional
spacetimes,\textquotedblright{} in \textit{Geometric Methods in Physics:
XXXII Workshop, Bia\l owie\.{z}a, Poland, June 30-July 6, 2013}, edited
by P. Kielanowski \textit{et al}. (Birkha\"user, 2014), pp. 233-242.

\bibitem{key-12}C. E. Wulfman, \textit{Dynamical Symmetry} (World
Scientific, 2011), p. 415.

\bibitem{key-13}W. I. Fushchich and A. G. Nikitin, \textit{Symmetries
of Maxwell's Equations} (D. Reidel, 1987), pp. 61-62.

\bibitem{key-14}W. I. Fushchich and A. G. Nikitin, \textit{Symmetries
of Equations of Quantum Mechanics}\\
 (Allerton Press, 1994), pp. 35-37.

\bibitem{key-15}Ref. \cite{key-4}, pp. 351-359.
\end{thebibliography}
\end{document}